\documentstyle[epsfig]{mn2e}

\def\gs{\mathrel{\raise0.35ex\hbox{$\scriptstyle >$}\kern-0.6em
\lower0.40ex\hbox{{$\scriptstyle \sim$}}}}
\def\ls{\mathrel{\raise0.35ex\hbox{$\scriptstyle <$}\kern-0.6em
\lower0.40ex\hbox{{$\scriptstyle \sim$}}}}

\begin{document}

\title
[Submm/far-IR SEDs] 
{Submillimetre and far-infrared spectral energy distributions of 
galaxies: the 
luminosity--temperature relation
and consequences 
for photometric redshifts} 
\author
[A.\,W. Blain, V.\,E. Barnard and S.\,C. Chapman]
{
A.\,W. Blain,$^{1}$ V.\,E. Barnard,$^{2}$ and S.\,C. Chapman$^{1}$
\vspace*{1mm}\\
$^1$ California Institute of Technology,
105-24, Pasadena, CA 91125, USA\\
$^2$ Cavendish Astrophysics, Madingley Road, Cambridge, CB3 0HE\\
}
\maketitle

\begin{abstract}
The spectral energy distributions (SEDs) of dusty high-redshift  
galaxies are poorly sampled in frequency and spatially unresolved. Their form 
is crucially important for estimating the large 
luminosities of these galaxies accurately, 
for providing circumstantial evidence concerning 
their power sources, and for estimating their redshifts 
in the absence of spectroscopic information. 
We discuss the suite of parameters necessary to describe their SEDs 
adequately without introducing unnecessary  
complexity. We compare directly four popular descriptions, explain 
the key degeneracies between the parameters in each  
when confronted with data, and highlight  
the differences in their best-fitting values. Using one representative 
SED model, 
we show that fitting to even a large number of radio, submillimetre and 
far-infrared (far-IR) continuum colours provides almost 
no power to discriminate between the redshift and dust 
temperature of an observed galaxy, unless  
an accurate relationship with a tight scatter exists 
between luminosity and temperature for the whole 
galaxy population.  
We review our knowledge of this 
luminosity--dust temperature relation derived from three
galaxy samples, 
to better understand the size of these uncertainties. Contrary to recent 
claims, we
stress that far-IR-based photometric redshifts are unlikely 
to be sufficiently accurate to impose useful constraints on 
models of galaxy evolution: finding  
spectroscopic redshifts for distant dusty 
galaxies will remain 
essential. 
\end{abstract}  

\begin{keywords}
radiation mechanisms: thermal -- dust, extinction -- galaxies: general -- 
galaxies: photometry -- infrared: galaxies -- submillimetre
\end{keywords}

\section{Introduction}

The rest-frame far-infrared (far-IR) 
thermal emission from dust grains heated by various 
sources -- 
the diffuse interstellar radiation field (ISRF) in galaxies, sites of active
star formation, and  
a central active galactic nucleus (AGN) -- can 
dominate the spectral energy distribution (SED) of galaxies 
(Soifer \& Neugebauer 1991; 
Sanders \& Mirabel 1996). The most luminous galaxy apparent in the 
Universe (APM\,08279+5255; Irwin et al.\ 1998) emits approximately 
60\,per cent 
of its bolometric luminosity in the far-IR waveband, while low-redshift 
galaxies with blue optical colours that were detected by 
the {\it IRAS} satellite also release about 60\,per cent 
of their total bolometric luminosity as  
thermal radiation from dust (Mazzarella \& Balzano 1986). Even the most 
quiescent spiral galaxies such as 
the Milky Way emit of order 30\,per cent of their 
total luminosity from dust (Reach et al.\ 1995; Alton et al.\ 1998; 
Dale et al.\ 2001; Dale \& Helou 2002). 
Dust emission remains important at high redshifts.  
The most distant quasi-stellar objects (QSOs) (Benford et al.\ 1999; 
Carilli et al.\ 2001; Isaak et al.\ 2002) and more typical,  
but still very luminous galaxies detected in submillimetre(submm) wave surveys
(Blain et al.\ 2002; Smail et al.\ 2002) emit 
strongly at rest-frame far-IR wavelengths. 

As compared with the rich variety of features in the SEDs of galaxies at 
near-IR, optical and ultraviolet 
wavelengths, the far-IR SED is simple, 
dominated by a smooth pseudo-thermal continuum emission spectrum. At most about 
1\,per cent of 
the emitted energy is associated with spectral lines from atomic fine-structure 
and molecular rotational transitions (Malhotra et al.\ 1997; 
Luhman et al.\ 1998; Combes, Maoli \& Omont 1999; Blain et al.\ 2000). The 
mid-IR spectra of galaxies from 10 to 30\,$\mu$m 
are expected to be significantly more 
complex, especially because of broad line emission from polycyclic aromatic 
hydrocarbon (PAH) molecules (Dale et al.\ 2001). 

A variety of models have been used to describe the far-IR SEDs of dusty 
galaxies. We compare four well-constrained descriptions 
with data for a variety of types of galaxy, and
highlight the importance both of degeneracies between the 
parameters and the need to avoid baroque descriptions that require a 
greater number 
of parameters than 
can be justified and fixed by existing data. 
Using one uniform, 
self-consistent description of the SED we 
discuss the accuracy of photometric redshifts that can be derived 
for high-redshift galaxies based on their observed colours, making assumptions 
concerning their 
SEDs. We describe in detail the degeneracy between redshift and dust temperature
when fitting photometric data for high-redshift galaxies 
(Blain 1999b; Blain et al.\ 2002), and discuss  
the prospects for 
breaking this degeneracy using information about absolute luminosity, 
obtained from a luminosity--temperature ({\it LT}) relation for dusty 
galaxies. A
narrow range of SEDs was 
included implicitly in recent 
discussions of the prospects for determining mm-wave 
photometric redshifts (Hughes et al.\ 2002; Aretxaga et al.\ 2003; 
Dunlop et al.\ 2003), which 
leads to encouraging results.  
We discuss existing data on the {\it LT} relation (Dunne et al.\ 2000; Stanford et 
al.\ 2000; 
Dale et al. 2001; Dale \& 
Helou 2002;  
Garrett\ 2002; Barnard \& Blain 2003; Chapman et al. 2003), which leads to 
a much less optimistic outlook for far-IR/submm photometric redshifts.  
The observed 
dispersion in the {\it LT} relation is the key quantity that limits the 
effectiveness of the technique. 

In Section 2 we describe 
four SED models, and compare them with a range of observed 
galaxy SEDs. We highlight the consequences of errors in 
the fitted SEDs and the {\it LT} relation 
for determining 
photometric redshifts in Section 3. 
Finally, in Section 4, 
we describe the requirements for spectroscopic observations that 
will remove this uncertainty, and describe the opportunities that 
much more detailed far-IR SEDs measured using 
{\it SIRTF}\footnote{See http://sirtf.caltech.edu} from 2003 will provide for 
better understanding the 
{\it LT} relation and for determining far-IR-based photometric redshifts. 

\section{SED descriptions} 

Various functions have been used to describe the quasi-blackbody 
far-IR/submm of the SEDs of 
dusty galaxies. The parameters that define the SED 
generally 
disguise the 
inevitably complex geometrical mix of dust grains at different temperatures 
in the interstellar medium of these galaxies, which are often disturbed and
interacting, and sometimes very luminous. The far-IR emission is 
visible at 
different optical depths in both emitted and scattered radiation. 

\subsection{Single-temperature models} 

The simplest SED description is based on a blackbody spectrum 
$B_\nu \propto \nu^3 / [ \exp(h\nu/kT) - 1 ]$ at a single temperature 
$T$, as a function of frequency $\nu$, modified by a 
frequency-dependent emissivity function $\epsilon_\nu \propto \nu^\beta$,   
where $\beta$ is in the range 1--2 (Hildebrand 1983). This 
yields an SED function is 
\begin{equation} 
f_\nu = \epsilon_\nu B_\nu \propto \nu^{3+\beta} / [ \exp(h\nu/kT) - 1 ].  
\end{equation}
Note that this function has an exponential Wien 
dependence when $\nu \gg kT/h$. It is necessary to modify this to 
a shallower form in order to agree with observed SEDs 
(see Fig.\,1). A straightforward way to counteract the mid-IR Wien tail is to 
substitute a power-law SED, $f_\nu \propto \nu^{-\alpha}$ at high 
frequencies, matching the power-law and thermal function (equation 1) 
with a smooth gradient at a frequency $\nu'$, which requires the condition 
${\rm d\,ln}f_\nu(\nu')/{\rm d\,ln}\nu' = -\alpha$ to be satisfied. 
Three parameters are required to describe the SED: $T$, $\beta$ and $\alpha$. 
The dust temperature $T$ determines the frequency of the SED 
peak, the emissivity index 
$\beta$ fixes 
the power-law index of the SED in the Rayleigh--Jeans regime, and $\alpha$ 
sets the slope of the mid-IR SED. This SED 
was used in the context of studying submm-wave 
galaxy evolution by Blain et al.\ (1999a), and has been used 
without the Wien correction  
to fit low-redshift SEDs by Dunne et al.\ (2000). 

An alternative `optically thick' functional form substitutes a more 
complex emissivity 
function, $\epsilon_\nu \propto [1 - \exp({\nu/\nu_0})^\beta]$, 
to describe the expected increase in the optical depth of dust emission at
higher frequencies, 
leading to 
an SED function, 
\begin{equation} 
f_\nu = \epsilon_\nu B_\nu \propto [1 - \exp({\nu/\nu_0})^\beta] B_\nu.
\end{equation}
This SED 
has been used by several authors, especially those 
dealing with the SEDs of galaxies and AGN at the highest redshifts (e.g. 
Benford et al.\ 1999; 
Isaak et al.\ 2002) where the SED is probed close to its rest-frame peak. This 
SED is identical to the $T$--$\alpha$--$\beta$ form at long wavelengths, but tends to 
a pure blackbody at frequencies greater than 
$\nu_0$, as   
expected from an optically thick source. This function also 
requires a power-law to temper the SED on the Wien tail, using the 
parameter $\alpha$. Four SED parameters are thus required  
in this model: 
$T$, $\alpha$ and $\beta$ as before, plus $\nu_0$. There is a strong 
degeneracy between the value of $\nu_{\rm 0}$ 
and the values of $T$ and $\beta$ (see Section\,2.4). 
Hence, a reasonable value of 
$\nu_{\rm 0}$ that  
corresponds to a frequency close to the 60- and 100-$\mu$m {\it IRAS} bands is 
usually assumed. Including this frequency-dependent opacity allows 
a more physical description of the SED, but the parameter 
$\nu_{\rm 0}$ is difficult to determine unambiguously from available observed 
data. 

\begin{figure*}
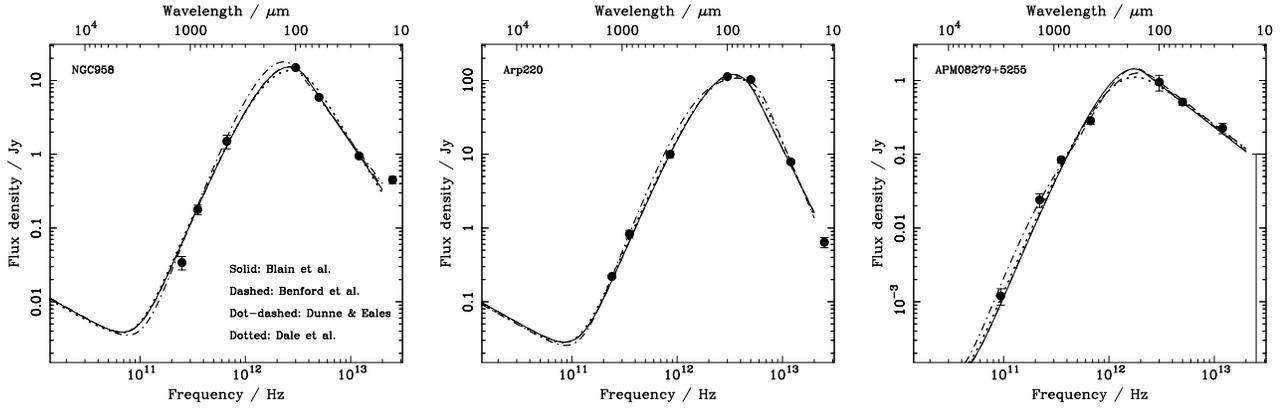

\begin{minipage}{170mm}
\begin{center}
\epsfig{file=SED_NGC958.ps, width=5.3cm, angle=-90} \hskip 3mm
\epsfig{file=SED_Arp220.ps, width=5.3cm, angle=-90} \hskip 3mm
\epsfig{file=SED_APM.ps, width=5.3cm, angle=-90}
\end{center}
\caption{The observed SEDs of three well-studied galaxies: the low-redshift 
($z=0.019$) Sb galaxy NGC\,958 (Dunne \& Eales 2001), the often-quoted 
prototype low-redshift ($z=0.018$) ultraluminous dusty galaxy 
Arp\,220, and the galaxy with the greatest 
apparent luminosity in the Universe APM\,08279+5255 (Irwin et al.\ 1998; 
Lewis et al.\ 1998). Four SED models described in 
Section\,2 are compared: a  
$T$--$\alpha$--$\beta$ model (equation 1), a model with a variable 
optical depth (equation 2), a model with both 
a cold and a warm dust 
component (equation 3) and a model with 
a power-law dust mass--temperature distribution (equation 4). 
The parameters required 
to fit the data in all four models are listed in Table\,1:
the numerical values differ, but 
all provide reasonable descriptions of the 
data, including the radio data for NGC\,958 and Arp\,220, which 
is not shown to avoid extending their abscissae over another 2 orders of 
magnitude. The plotted ranges of frequency are equal, 
demonstrating the range of different 
apparent dust temperatures/rest-frame SED peak frequencies 
and mid-IR spectral indices observed.}
\end{minipage}
\end{figure*} 

\begin{table*}
\caption{Lists of the best-fitting parameters in the four 
SED models described in Section\,2 (equations 1--4) 
required to reproduce the observed 
SEDs of three well-studied galaxies: see Fig.\,1. 
The bolometric luminosities $L$, defined as the integral under the 
SED $f_\nu$, normalized
to the observed flux density of the galaxy,  
associated with each model and galaxy are 
listed. In model 3 a cool component with 
a dust temperature $T_{\rm c}=20$\,K is always added. In model\ 4 the 
value of the upper limit to the dust temperature $T_{\rm max}=2000$\,K 
in all cases.}
\begin{center}
\begin{tabular}{lllll}
\hline 
Galaxy & Model 1 & Model 2 & Model 3 & Model 4  
\\
\hline
NGC\,958 & $T=28.8\pm1$\,K & $T=33\pm2$\,K & $T_{\rm w}=24.8\pm2$\,K & 
$T_{\rm min}=22.0\pm1$\,K \\
$z=0.019$ & $\alpha = 2.02 \pm 0.2$ & $\alpha = 2.0 \pm 0.2$  
& $\alpha = 1.9 \pm 0.2$  
& $\gamma = 7.9 \pm 0.3$ \\
	& $\beta = 1.5^f$ & $\beta = 1.5 \pm 0.1$ & $\beta = 2.0^f$ & 
$\beta = 1.5^f$ \\
	&  & $\nu_{\rm 0} = (2.9 \pm 0.5) \times 10^{12}$\,Hz & 
$F_{\rm wc} = 0.58 \pm 0.2$ &  \\
& $L = 3.1 \times {10^{11}}$\,L$_\odot$ & 
$L = 1.8 \times {10^{11}}$\,L$_\odot$ & 
$L = 2.6 \times {10^{11}}$\,L$_\odot$ & $L = 2.9 \times {10^{11}}$\,L$_\odot$\\ 
\noalign{\smallskip} 
Arp\,220 & $T = 37.4 \pm 1$\,K & $T=56\pm 1.5$\,K & $T_{\rm w}=42.3 \pm 1$\,K 
& $T_{\rm min}=29.7 \pm 1$\,K \\
$z=0.018$ & $\alpha = 2.9 \pm 0.2$ & $\alpha = 3.0 \pm 0.1$ & 
$\alpha = 3.43 \pm 0.3$ & $\gamma = 8.8 \pm 0.2$ \\
& $\beta=1.5^f$ & $\beta=1.55 \pm 0.1$ & $\beta=2.0^f$ & $\beta = 1.5^f$ \\
	& & $\nu_{\rm 0} = (1.46 \pm 0.1) \times 10^{12}$\,Hz  
& $F_{\rm wc} = 0.51 \pm 0.1$ & \\
 & $L=1.41 \times {10^{12}}$\,L$_\odot$ & $L=1.43 \times {10^{12}}$\,L$_\odot$ 
& $L=1.47 \times {10^{12}}$\,L$_\odot$
& $L=1.39 \times {10^{12}}$\,L$_\odot$ \\
\noalign{\smallskip}
APM\,08279+5255 & $T = 91 \pm 5$\,K & $T=187\pm 10$\,K & 
$T_{\rm w}=83 \pm 3$\,K & 
$T_{\rm min}=59.7 \pm 3$\,K \\
$z=3.8$ & $\alpha = 1.1 \pm 0.1$ & $\alpha = 1.05 \pm 0.15$ &
$\alpha = 1.1^f$ & $\gamma = 6.5 \pm 0.2$ \\
& $\beta=1.5^f$ & $\beta=1.7 \pm 0.3$ & $\beta=2.0^f$ & $\beta = 1.5^f$ \\
        & & $\nu_{\rm 0} = (1.4 \pm 0.2) \times 10^{12}$\,Hz 
& $F_{\rm wc} = 0.033 \pm 0.01$ & \\
 & $L=6.7 \times {10^{15}}$\,L$_\odot$ & $L=3.2 \times {10^{15}}$\,L$_\odot$ & 
$L=1.0 \times {10^{16}}$\,L$_\odot$ & $L=3.1 \times {10^{15}}$\,L$_\odot$ \\
\hline 
\end{tabular}
\end{center}
$^f$ This parameter value was fixed in the fitting process, either 
to reduce the size of the parameter space being 
searched or to enforce a physically meaningful value. 
\end{table*}

\begin{figure}
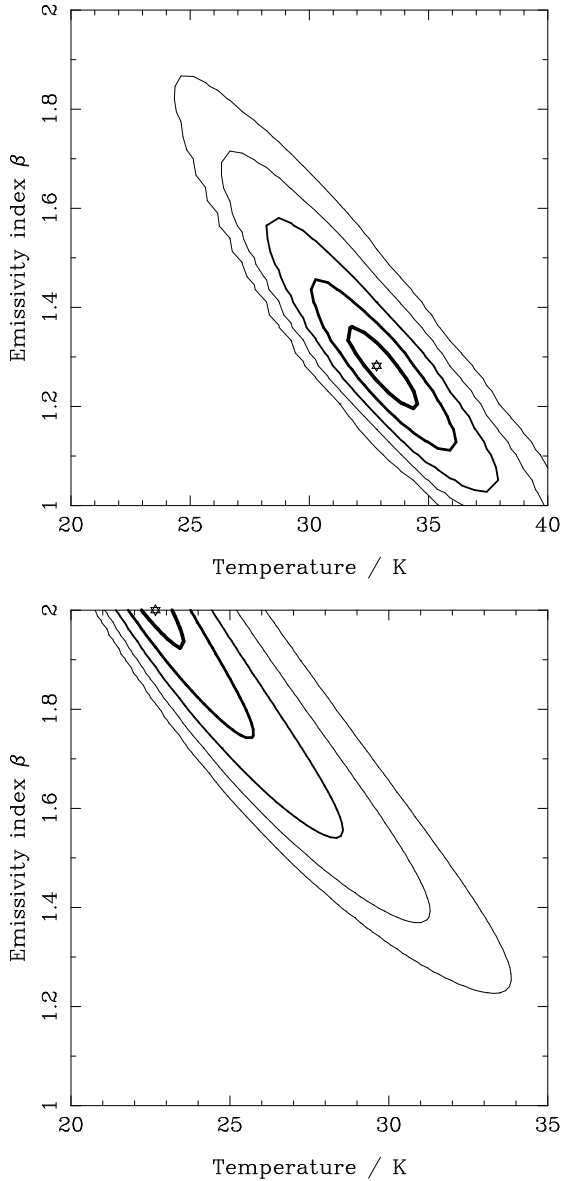

\begin{center}
\epsfig{file=958_Tbeta.ps, width=7.65cm, angle=-90} \vskip 3mm
\epsfig{file=958_Tbeta_mm.ps, width=7.65cm, angle=-90} 
\end{center} 
\caption{ 
Contours of the probability of fitting data for NGC\,958,
using model 1, as a function of the values of the most 
important model parameters $T$ and $\beta$. The result for 
all the data plotted in Fig.\,1, 
plus radio data, is shown in the upper panel, while 
in the lower panel the results are derived for only the subset of data 
at 850, 450, 100 and 60\,$\mu$m 
for consistency with 
Dunne \& Eales (2001). 
The contours are 
spaced by unit standard deviations away from the best-fitting 
values, which are marked by a star. A fixed 
best-fitting value of $\alpha = 2.02$ is assumed, 
and $\beta$ spans 
the physically plausible range $1 \rightarrow 2$. There is a  
strong 
degeneracy between $T$ and $\beta$ in both panels. 
Between the two panels the best-fit points remain on a similar 
$T$--$\beta$ trend line, but lie on different tracks. The best-fitting 
point from the restricted data set prefers the 
physically implausible $\beta > 2$ region. Note that if $\beta$ is 
fixed at 1.5, then the value of $T=28.8$\,K listed in Table\,1 is the 
best fitting value in the upper panel.}
\end{figure} 

\begin{figure}
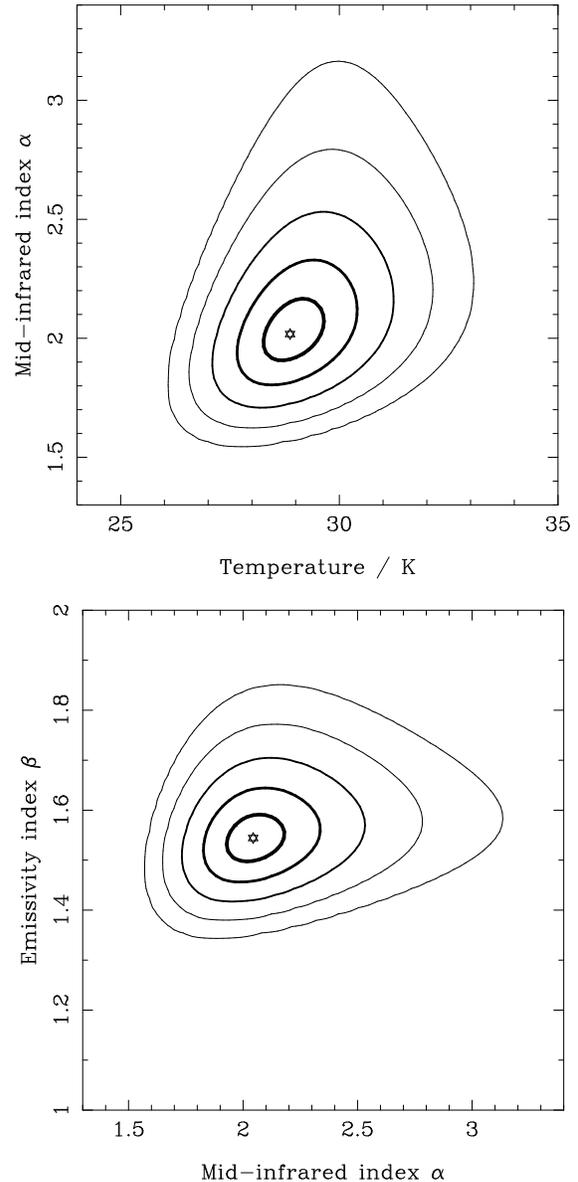
 
\begin{center} 
\epsfig{file=958_Talpha.ps, width=7.65cm, angle=-90} \vskip 3mm
\epsfig{file=958_ab.ps, width=7.65cm, angle=-90} 
\end{center}
\caption{The counterpart to Fig.\,2 for the two other 
pairs of parameters: $T$--$\alpha$
(upper panel; with $\beta = 1.5$)  
and $\alpha$--$\beta$ (lower panel; with $T=29$\,K). 
In both panels the best-fitting values  
lie in a well-defined, circular region of the parameter
space.}
\end{figure}

\subsection{Models with multiple dust temperatures} 

Descriptions of the 
SED can include more than one
dust temperature. Most notably, these include models based on 
radiative transfer calculations, 
in which a continuous distribution of sources is 
assumed in 
some geometry, and the temperature distribution 
of the dust as a function of position is calculated
self-consistently to build up an SED 
(Granato, Danese \& Franceschini 
1996; Devriendt et al. 1999; Efstathiou, Rowan-Robinson 
\& Siebenmorgen 2000). Note, 
however, that 
even for nearby galaxies the spatial and spectral resolution available is 
insufficient to constrain the $\sim 10$ parameters required to 
describe this type of model even in the simplest spherical geometry.  
When the sub-arcsec 
resolution of the Atacama Large Millimeter Array (ALMA) 
interferometer and {\it James Webb (Next Generation) Space 
Telescope (NGST)} are available at submm and near/mid-IR 
wavelengths, respectively, then radiative transfer models 
will have a role to play in interpreting observations. At present, 
the quality of available data does not justify the incorporation of such 
complexity. 

Two more practical multi temperature SED descriptions have been used. 
A two-temperature model by Dunne \& Eales (2001) includes a cool component 
at a fixed temperature of $T_{\rm c}=20$\,K to describe dust heated by 
the general diffuse ISRF 
of the galaxy, and a component of hotter dust at a temperature 
$T_{\rm w}$, with a 
mass fraction $F_{\rm wc}$, that is 
heated more intensely in star-forming regions. 
Each component is described by a 
modified blackbody $\nu^\beta B_\nu$ spectrum, assuming a fixed value of the 
emissivity index $\beta=2$. The resulting SED function  
\begin{equation} 
f_\nu \propto \nu^\beta [ B_\nu(T_{\rm c}) + F_{\rm wc} B_\nu(T_{\rm w})].
\end{equation}
There are 3 free parameters in this model 
if $T_{\rm c}$ is fixed at 20\,K and $\beta$ is fixed at 2.0: 
$T_{\rm w}$ and $F_{\rm wc}$ to describe the thermal part of the SED, and 
a power-law index $\alpha$, to fix 
the mid-IR SED. A similar two-temperature model, with a greater spread 
between the temperatures, is described in the context of the dust emission 
from galaxies that are members of the Virgo cluster by Popescu et al.\ 
(2002). 

A fourth, physically motivated, and yet still adequately constrained 
model was described by Dale et al.\ (2001), who assumed  
a power-law distribution of dust masses as a function of 
temperature, in which the mass of dust heated to a temperature between 
$T$ and $T + {\rm d}T$, is given by $m(T) \propto T^{-\gamma}$. The spectral 
contribution to 
the SED from each temperature component is $\nu^\beta B_\nu$, and so the 
composite SED is given by the integral 
\begin{equation} 
f_\nu \propto \int_{T_{\rm min}}^{T_{\rm max}} m(T) \nu^\beta B_\nu 
\, {\rm d}T \propto 
\int_{T_{\rm min}}^{T_{\rm max}} T^{-\gamma} \nu^\beta B_\nu 
\, {\rm d}T.  
\end{equation} 
The value of $\gamma$ effectively determines the mid-IR slope 
at frequencies
$\nu \gg kT/h$, and 
has a close equivalence to $\alpha$ above. Thus 
there is no 
need to introduce another parameter to counteract the Wien tail of the SED 
here. The 
value of $\gamma$ required to produce the 
same mid-IR spectrum as the other three models is 
$\gamma \simeq 4 + \alpha + \beta$ 
(Blain 1999a). The role of $T_{\rm min}$ is 
equivalent to $T$ in the other models, and determines the frequency of the 
peak of the SED, subject to 
a weak dependence on the value of $\gamma$. $T_{\rm min}$ must 
always exceed the  
cosmic microwave background (CMB) temperature. 
The value of the maximum temperature $T_{\rm max}$ is 
relatively unimportant, unless $T_{\rm min}$ is very high: 
$T_{\rm max}$ 
was always set to 2000\,K to represent the sublimation temperature of the 
dust. We have thus chosen to keep $T_{\rm max}$ fixed and vary both 
$T_{\rm min}$ and $\gamma$ to fit the data. 
This model has the minor practical disadvantage that an integral must be 
performed, or a look-up table employed, to evaluate the SED. In the 
updated model of Dale \& Helou (2002) the value of the emissivity 
parameter $\beta$ at wavelengths longward of 100\,$\mu$m is now 
parametrized as a function of the intensity of the ISRF. 
For simplicity, we adopt a constant value $\beta = 1.5$ here. 

For all four models an additional component of synchrotron radio emission was 
added, by assuming the conventional far-IR--radio correlation (Condon 1992) 
between the 1.4-GHz radio  
emission and the flux densities in the 60- and 100-$\mu$m {\it IRAS}
passbands. This correlation holds 
with a 0.2-dek dispersion 
over 4 orders of magnitude in luminosity, and 
should provide a 
reasonable representation of the expected radio flux in the absence of 
additional radio emission from electrons accelerated by an AGN. 
The details of 
the extrapolation method are described in Blain (1999a). 

\subsection{Comparison of the SED models with data} 

The four SED models described above were fitted to data 
for three well-studied galaxies with very different luminosities and 
rest-frame SEDs. The results are shown in Fig.\,1 and Table\,1. 
The data were obtained from  
{\it IRAS} between 
12 and 100\,$\mu$m, from SCUBA at 450 and 850\,$\mu$m, from other 
ground-based 
mm-wave telescopes, and from the VLA at 
1.4\,GHz in the cases of NGC\,958 and Arp\,220. The SEDs are 
reasonably well sampled all the way from 
the mid-IR to the radio waveband. All four 
models can provide a good description of these SEDs. In models 1 and 4,  
the emissivity index $\beta$ was 
fixed to 1.5 to reduce the size of the parameter space to search in the 
maximum-likelihood 
fitting routine, without 
great loss of generality or hampering the quality of the fit. A key degeneracy
between fitted values of $T$ ($T_{\rm min}$ in model\,4) 
and $\beta$ is described in the next section. 

The width of the peak of the model SEDs, defined as the fractional 
frequency range over which the SED is reduced to half of its peak value, 
is the feature that 
differs most significantly 
from model to model, but never does so by more than 
a factor of approximately 1.5. The different functional forms of the SED  
require values of temperature that can have a large dispersion. 
This is 
especially true 
for APM\,08279+5255. In model\,2 the 
dust cloud is inferred to become optically thick 
at frequencies less than 
the peak of the SED, requiring a much higher temperature to describe 
the data than in the 
optically thin models, reflecting the lower effective 
value of $\beta$ close to the peak of the SED in the optically-thick
model. 
The peak frequency of the fitted SED is consistent with the data, 
and the luminosity is within the range spanned by the 
other three models (see Fig.\,1 and Table\,1). Hence, model\,3 
still provides an adequate description of the data. 
For the other galaxies, 
the temperature required to fit the data in model\,1 
is systematically less 
than that in the optically-thick model\,2, again 
because of the effectively  
smaller value of $\beta$ at the SED peak in  
model 2. 

The temperature of the warm component required to fit SEDs 
in the two-temperature 
model\,3 depends on the relative intensities of the 25- and 100-$\mu$m 
{\it IRAS} flux densities. The value of $T_{\rm min}$ 
required to fit the data in model\,4 with a
power-law dust mass distribution is always lower than in the 
other models, reflecting its 
definition  
as a lower limit to a distribution of hotter  
temperatures. The factor by which it is lower than in the other 
models depends on the value of the mass--temperature function index 
$\gamma$ (equation 4): a steeper decline in the proportion of dust 
at higher temperatures 
leads to a smaller difference, while a greater fraction of hot dust 
corresponds to a greater difference. 

\subsection{Degeneracies between fitted SED parameters} 

The most significant practical degeneracy in fitting the results shown in 
Fig.\,1 in all four models 
is between the dust temperature $T$ and the emissivity power-law 
index $\beta$, as illustrated in Fig.\,2. This occurs because the 
peak frequency of the SED scales approximately with the value of 
$\beta / T$, and so the ratio of $\beta$ and $T$ must remain approximately 
constant in order to reproduce the data. In model\,4 there is an 
effective dust temperature $T$ that reflects the range of temperatures 
present. For a fixed value of $\gamma$ this effective temperature is 
determined by the value of $T_{\rm min}$. 

Considering a 
subset of the data can modify the position of the best fit values significantly 
along the extended direction of the probability contours in the figure. 
The other pairs of parameters $T$--$\alpha$ and $\alpha$--$\beta$ do not show 
such a degeneracy -- the probability contours determined for their fit to 
the SED data are almost circular (Fig.\,3) -- and so these pairs of 
parameters have unique well-determined values when the model provides  
a good description of the observed SED. 
Note that the bolometric far-IR luminosity of the galaxy that 
is derived by integrating the SED in frequency from 100\,GHz to the 
frequency equivalent to a wavelength 
of 1\,$\mu$m changes 
along the ridge of the probability 
curve shown in the upper panel of Fig.\,2. The inferred luminosity 
increases smoothly along the ridge from 
$2.6 \times 10^{11}$\,L$_\odot$ when $T=23$\,K to 
$4.4 \times 10^{11}$\,L$_\odot$ when $T=40$\,K, with 
$L = 3.1 \times 10^{11}$\,L$_\odot$ at the best-fitting value of 28\,K 
if $\beta=1.5$. Hence, 
there is little practical difficulty in using this 
description -- neither the luminosity nor the peak frequency of the 
SED differs significantly as the permitted region in the $T$--$\beta$ 
parameter space is traversed. There is however a real problem in trying to 
associate the values of the fitted parameters with the true 
physical properties 
of the dust grains that generate the emission. 

With the exception of $T$ and $\beta$ in all models ($T_{\rm min}$ and 
$\beta$ in model\,4) and 
the pairs $T$--$\nu_0$ and
$F_{\rm wc}$--$T_{\rm w}$
in models\,2 and 3 respectively, 
the parameters are 
well determined. The well-determined pairs of parameters $T$--$\alpha$ and 
$\alpha$--$\beta$ are illustrated in Fig.\,3, with almost circular 
probability contours. The probabilities for the 
relatively ill-constrained pairs discussed above are illustrated in Figs\,
4 and 5. 
There are two points to note from these probability figures. 
First, along the 
track of maximum probability in these figures, the value of the 
associated bolometric far-IR  
luminosity of the galaxy changes by only 5--10\,per cent. This is 
much less variation than the factor of 2 variation across the 
wider range of the $T$--$\beta$ parameter space shown in Fig.\,2. Secondly, 
$F_{\rm wc}$ is not defined accurately  
even by the excellent data for NGC\,958, as shown in Fig.\,5. 
Hence, this is likely to 
provide the least informative presentation of SED data 
amongst the four models used. 
For APM\,08279+5255, the extra population of 20-K dust in this 
model can be seen generating 
a break in the low-frequency slope of the SED in Fig.\,1. 
The mass of cold dust present in far-IR luminous galaxies 
may dominate the total mass of dust at all temperatures, and  
provide information concerning the history of metal enrichment within, but  
it is far from energetically dominant, and difficult to measure to better 
than a factor of a few. 

\begin{figure}
\begin{center}
\epsfig{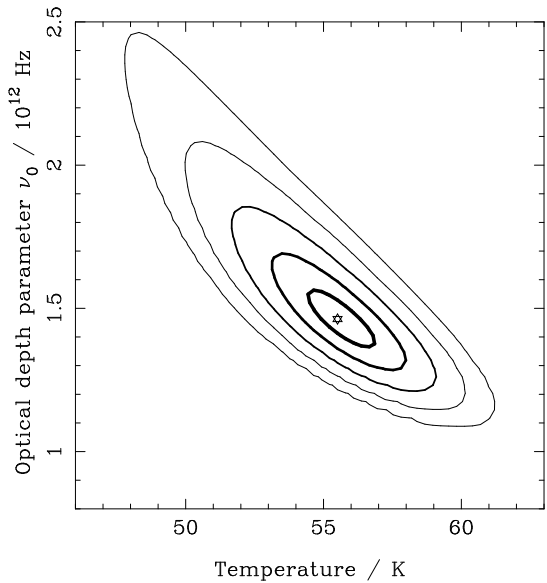}
\end{center}
\caption{The counterpart to Fig.\,2, for the optically thick SED
(model\,2) fitted to data for Arp\,220. 
The shape of the probability contours illustrates the 
degeneracy between temperature and the frequency at which a 
galaxy becomes optically thick, $\nu_0$.
$\alpha=3.0$ and $\beta=1.5$ are assumed, the best-fitting values. 
}
\end{figure} 

\begin{figure}
\begin{center}
\epsfig{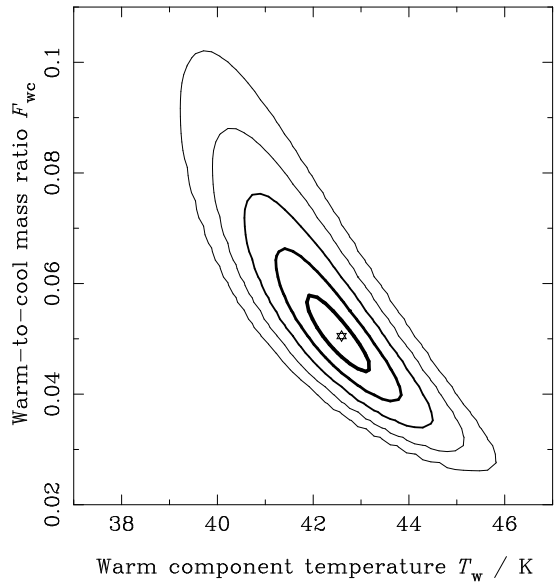}
\end{center}
\caption{An illustration of the significant degeneracy between the warm dust 
temperature parameter $T_{\rm w}$ and the warm dust mass fraction 
parameter $F_{\rm wc}$ obtained from fitting to the SED data for 
Arp\,220, at a known redshift $z=0.018$ using 
the two-temperature model 3. 
The large extent of the error ellipse in the direction of 
$F_{\rm wc}$ reflects the 
general difficulty of determining a dust mass from SED data, 
illustrating the difficulty of 
associating parameter values in an SED fit with the true 
physical properties 
of dust grains in an  
observed galaxy.
} 
\end{figure} 

For consistency with our earlier 
treatments (Blain et al.\ 1999a, 2002; Barnard 
\& Blain 2003) we will adopt the $T$--$\alpha$--$\beta$ model (model\,1) in 
the following discussions. Subject to the effective dust temperature 
$T$ being slightly different from that in the other SED parametrizations, 
this model provides a good description of   
observations for galaxies ranging from low-luminosity 
spirals to the most luminous high-redshift systems (Fig.\,1). The results 
that follow are not only valid for this description of the SED, but 
are generic results that apply to all 4 descriptions discussed above. 

\begin{figure}
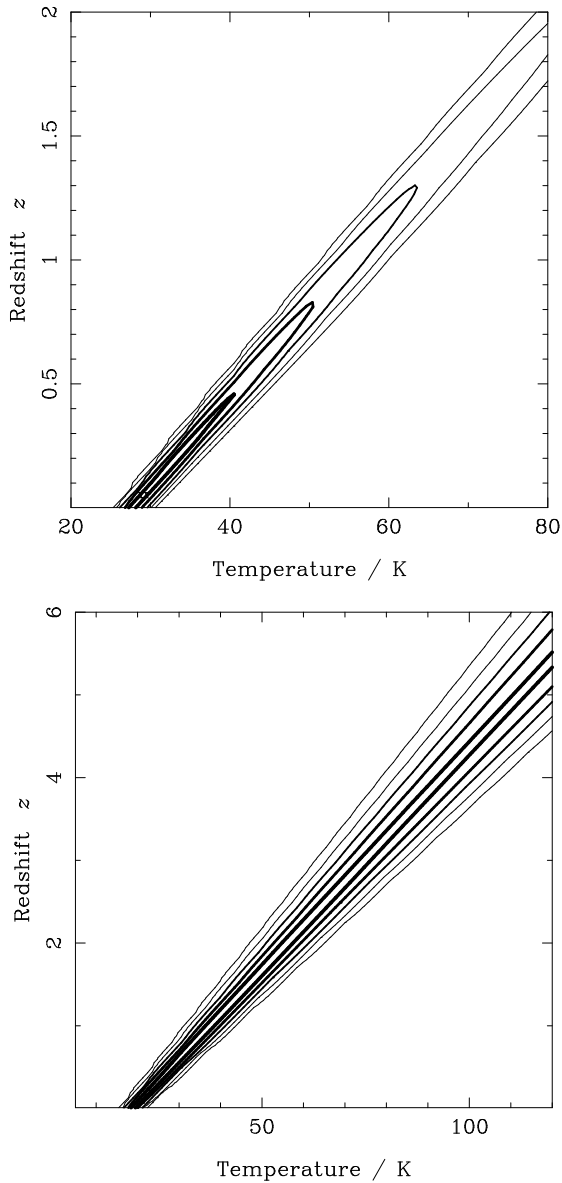

\begin{center}
\epsfig{file=958_Tz.ps, width=7.65cm, angle=-90} \vskip 3mm
\epsfig{file=APM_Tz.ps, width=7.65cm, angle=-90}
\end{center} 
\caption{An illustration of the degeneracy between the dust temperature 
$T$ and the redshift $z$ fitted to SED data for two 
dusty galaxies in Fig.\,1, disregarding their known redshifts: 
NGC\,958 at $z=0.019$ (upper panel) 
and APM\,08279+5255 at $z=3.8$ 
(lower panel).  
The $T$--$\alpha$--$\beta$ model\,1 SED is assumed, with $\alpha$ and 
$\beta$ fixed at their values from Table\,1 
to minimize the scatter in the fitted values. 
The data define a narrow 
track aligned with the locus of constant $T/(1+z)$.
The radio data for NGC\,958 
leads to the reduction in 
probability from low to high temperatures along the track of the 
contours in the upper panel. 
Observations of colours alone, even 
with radio data, provide a strong constraint only on the ratio  
$T/(1+z)$.} 
\end{figure} 

\section{Photometric Redshifts}  

The well-defined pseudo-thermal SED of dusty galaxies (Fig.\,1) offers 
a prospect of recognizing the redshifts of galaxies with the 
same intrinsic SEDs by comparing far-IR and
submm colours. This was discussed in the context of 
identifying 
high-redshift galaxies amongst more numerous low-redshift galaxies in a 
shallow submm-wave survey 
by Blain (1998) and for observations  
of the first generation of hard-to-identify submm galaxies, with 
flux-density limits  
from {\it IRAS} observations
by Hughes et al.\ (1998) and 
Eales et al.\ (1999). Eales et al.\ noted that it is 
not possible, a priori, to be certain of whether 
a dusty galaxy is hot and far away, or cool and close by, and that there could 
be significant consequences for the cosmological implications of 
the population of submm-luminous galaxies if the 
dust temperature/redshift is not estimated correctly. If the 
dust temperature 
defining the SED is too hot, and/or 
the redshift of the population is too 
great, then the cosmological importance of 
submm
galaxies can be overstated. 
The reason for the similar effects of increasing both redshift 
and temperature is that the peak of the SED is determined by the value 
of $\nu/T$ in the exponential term of the Planck function. Redshifting 
the spectrum by a factor of $(1+z)$ in frequency $\nu$ thus 
has a directly equivalent effect 
to modifying the temperature $T$ by the same fraction. 
 
By assuming a narrow range of  
SED templates, with a tight distribution of dust temperatures, 
when trying to match a redshift (Hughes et al.\ 
2002; Dunlop et al.\ 2003),  
this large degeneracy can appear to vanish, leading to 
unrealistically optimistic estimates of the accuracy of the derived 
redshifts  
to $\Delta z \simeq 0.5$. Unless a representative 
range of SED templates is available, perhaps including the full range of 
observed SEDs of dusty galaxies, with temperatures 
from less than 20\,K (Reach et al.\ 1995) 
to more than 80\,K (Table\,1), then the errors on photometric redshifts
could be underestimated. The true error is at 
least as great as 
the fractional uncertainty in the dust temperature, even if there are
very small errors on the photometric data itself. 

This dust temperature--redshift degeneracy is illustrated in Fig.\,6 for 
the photometric data available for both the low-redshift low-luminosity 
galaxy NGC\,958, which includes radio data, and  
the high-redshift, high-luminosity dusty QSO APM\,08279+5255. 
The availability of 
radio data reduces the degeneracy somewhat, as the different emission 
mechanisms 
lead to different spectral indices (Carilli \& Yun 1999; Yun \& Carilli 2002); 
however, 
the radio--submm colour is still 
a much better indicator of the ratio $T/(1+z)$ than of 
$T$ and $z$ separately (Blain 1999a). The degeneracy lies along the 
locus $T \propto (1+z)$: see Fig.\,6. If a fraction of the radio emission is 
generated by an AGN, then the derived redshift will be underestimated by 
perhaps a large amount. 

A practical example of a high-redshift galaxy for which 
photometric redshifts may be sought is SMM\,J14011+0252, with a known 
redshift $z = 2.56$ (Frayer et al.\ 1999), and a well 
determined radio--far-IR SED
(Ivison et al.\ 2001). The galaxy has an achromatic gravitational 
lensing magnification of a factor of 2.5.  
Upper limits from {\it IRAS} provide weak constraints on 
its dust temperature and mid-IR spectral index $\alpha$. 
Values of $T$ and $z$ that provide good fits to the SED data are shown 
in Fig.\,7, setting aside the known redshift $z=2.56$. 
The true redshift corresponds 
to $T \simeq 33$\,K; note, however, that the extent of the 
1$\sigma$ probability contour in the fit, even in the lower panel 
for which the assumed 
fractional errors on the photometric data points are 
reduced to an artificially 
low 2\,per cent, is $T = 29^{+14}_{-9}$\,K 
or $z = 2.2 \pm 1.2$, hardly  
useful redshift information. 
Without increasing the accuracy of the observed data, 
no useful result for redshift $z$ alone can be 
quoted, as the probability contours are open in the upper panel. 

\subsection{A luminosity--temperature ({\it LT}) relation to the rescue?}  

The fit for SMM\,J14011+0252, even with greater 
observational accuracy assumed, shows the large 
degeneracy between temperature and redshift when fitting photometric 
data. 
Is it possible to improve 
the redshift accuracy by inferring the luminosity for the 
galaxy if $T$ and $z$ are taken to lie along the  
tracks of maximum probability in the direction of $T=(1+z)$ 
in Figs\,5--7? If there is a known link between 
luminosity and dust temperature, an {\it LT} relation (Dunne et al.\ 2000; 
Dale et al.\ 
2001; Blain et al.\ 2002; Dale \& Helou 2002; Barnard \& Blain\ 2003; 
Chapman et al.\ 2003), 
then a colour-magnitude diagram 
could be used to locate galaxies on the degenerate $T=(1+z)$ tracks,
allowing redshifts to be determined. 
The inclusion of luminosity information is an 
implicit assumption in the photometric redshift technique 
with a narrow range of SED parameters discussed by Hughes et al.\ (2002), 
Aretxaga et al.\ 
(2003) and Dunlop et al.\ (2003).

\begin{figure}
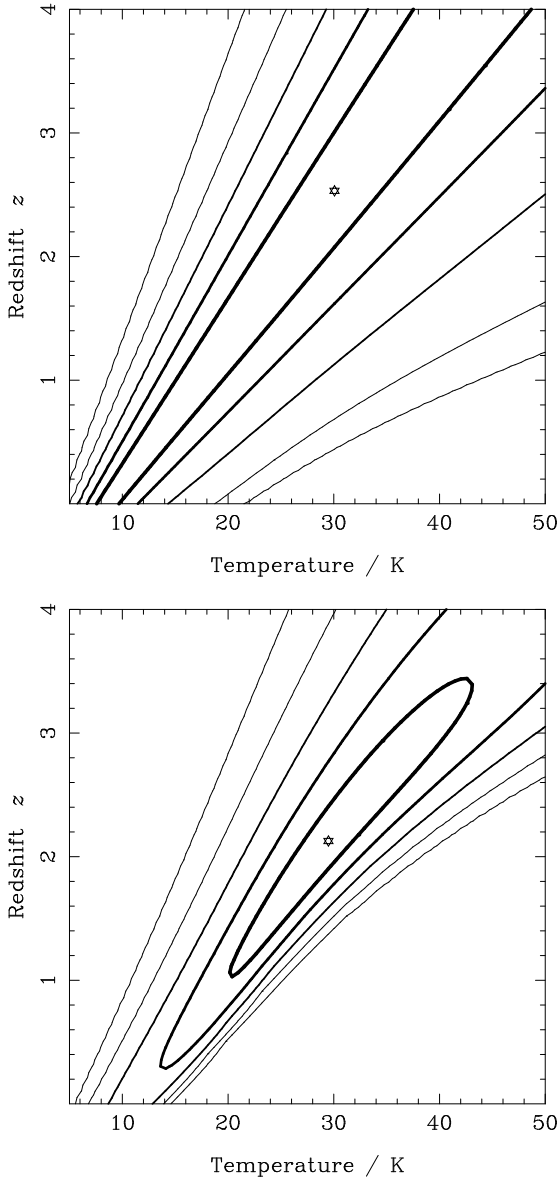
 
\begin{center} 
\epsfig{file=blobii_rad.ps, width=7.65cm, angle=-90} \vskip 3mm 
\epsfig{file=blobii_subrad2percent.ps, width=7.65cm, angle=-90} 
\end{center} 
\caption{A fit to temperature and redshift derived from all existing data 
for the $z=2.56$ 
submm galaxy SMM\,J14011+0252 (Frayer et al.\, 1999; Ivison 
et al.\ 2001). In the upper panel the true observational 
errors are assumed, while in the lower panel the fractional errors 
are set to only 2\,per 
cent, about five times better than the real data. In both panels there 
is a huge degeneracy in the direction $T \propto 1+z$, 
and so little meaningful redshift information is available 
from the colours alone, even with unreasonably small errors. 
In the lower panel there is a clear maximum 
probability. The best-fitting line is also 
deflected between low and high temperatures 
due to the different temperature dependence of the submm 
and radio SEDs. 
}
\end{figure} 

\begin{figure}
\begin{center}
\epsfig{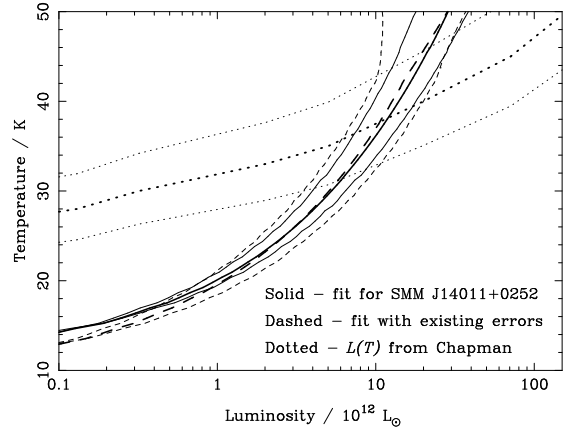}
\end{center}
\caption{The bolometric luminosity inferred from photometric data 
for SMM\,J14011+0252, 
as a function of temperature in the restframe of the galaxy along the 
permitted track of maximum probability in the $T$--$z$ parameter space 
as shown in Fig.\,7. 
The results derived assuming 
both 
existing errors (dashed lines) and tighter 2\,per cent errors (solid 
lines), are similar. The thin lines enclosing the thick lines trace 
out the region of the figure occupied by {\it LT} values within one standard 
deviation of the best-fit track through each part of Fig.\,7. 
In both cases the luminosity is corrected 
for the known magnification factor of 2.5. 
Each temperature corresponds to a different redshift $T \propto (1+z)$. 
At the known redshift $z=2.56$ the fitted temperature is 
approximately 35\,K (Fig.\,6). The low-redshift {\it LT} relation 
derived from {\it IRAS} data and its 0.14-dex interquartile range 
uncertainty 
(Dale et al.\ 2001; Dale \& Helou 2002; Chapman et al. 2003)
is shown 
by the thick and thin dotted lines respectively. The uncertainty in the 
{\it LT} relation 
dominates the  
error on an inferred $T$--$z$ value. The scatter in temperature $T$ by 
approximately 15\,per cent corresponds to a scatter in inferred $z$ by 
approximately 25\,per cent. 
}
\end{figure}  

\begin{figure}
\begin{center}
\epsfig{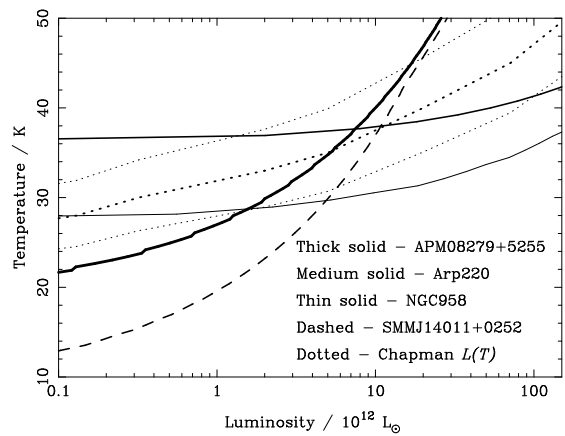}
\end{center}
\caption{The bolometric luminosity inferred from photometric data 
for the $z=0.019$ spiral galaxy NGC\,958, 
the $z=0.018$ ULIRG Arp220 
and the ultraluminous $z=3.87$ galaxy APM\,08279+5255 (demagnified 
by a factor of 50), compared 
with the result for SMM\,J14011+0252
and the low-redshift {\it LT} relation shown in 
Fig.\,8. The errors on 
the track traced for all three galaxies are very small.  
The curves for NGC\,958 and Arp220 intersect the 
{\it LT} relation close to their 
expected temperatures, but the curve for APM\,08279+5255 
does not. 
}
\end{figure}  

In Fig.\,8 we show the bolometric luminosity $L$ inferred for each temperature 
$T$ and redshift $z$ 
along the track 
of maximum probability in both the upper and lower panels of Fig.\,7. The 
luminosity is plotted as a function of temperature, but the unique redshift  
associated with each temperature 
can be read from the tracks in Fig.\,7. 
The ranges of fitted 
{\it LT} values that lie within 1-$\sigma$ of the most probable track in Fig.\,7 
are also shown enclosed by the thinner solid and dashed lines in Fig.\,8. 
The inferred luminosity increases with increasing 
temperature/redshift, and does so more rapidly than the  
{\it LT} relation inferred for low-redshift {\it IRAS} galaxies by 
Chapman et al.\ (2003), which is shown by the dotted lines in Fig.\,8 and 
has a scatter of 0.14\,dek in the 
interquartile range.  

By comparing the two panels of Fig.\,7 and the form of the {\it LT} curves for 
each panel shown in 
Fig.\,8 it is clear that reducing the size of the errors on the 
photometric 
data does not change the width of the inferred 
track in the luminosity--temperature/redshift space
significantly, whereas 
including luminosity information does restrict 
the range of plausible temperatures/redshifts for galaxies 
with photometric data to those lying between the thin dotted lines. 
Note, however, that the 1$\sigma$ spread in the low-redshift 
{\it LT} relation covers a range $T = 38 \pm 10$\,K, a dispersion of 
about 25\,per cent. The associated redshift, obtained by reading off the 
maximum probability track in Fig.\,7, 
based on unrealistically accurate 
data, indicates a range $z = 2.9 \pm 0.9$: the result for less precise 
existing data shown in the upper panel of Fig.\,7 is 
$z = 3.4 \pm 1.1$. While these ranges include the true redshift 
$z=2.56$, the 
uncertainties make the results of little use when compared to the 
exact values of $T$ and $L$ that could be found from a 
spectroscopic redshift. The spread in the derived redshifts, assuming the 
measured scatter in the low-redshift {\it LT} relation (Chapman et al.\ 2003) 
is a factor of 2 
greater than 
the photometric redshift accuracy claimed from the SMM\,J14011+0252 data 
by Hughes et al.\ (2002; $z=2.9^{+0.6}_{-0.4}$).  
The corresponding range in 
luminosity covers $6 \times 10^{12}$ to 
$2 \times 10^{13}$\,L$_\odot$, a factor of approximately 3 (Fig.\,8). 
Photometric redshifts cannot be obtained to
the accuracy proposed by Hughes et al.\ (2002) without assuming an  
unreasonably tight dispersion in the {\it LT} 
relation, by taking 
an inadequate range of SED functions into account. 

Knowledge of an {\it LT} relation for submm galaxies is thus 
unlikely to rescue far-IR/submm/radio photometric redshifts from 
the temperature--redshift degeneracy. The error in the derived 
temperature/redshift for an individual galaxy is expected to be dominated 
by the uncertainty in the {\it LT} relation. 
Once a spectroscopic redshift is obtained for a submm galaxy, 
however, its temperature and 
luminosity can be determined quite accurately: compare 
the widths of the probability contours in Figs\,5--7 
at constant redshift. 
Temperatures and luminosities accurate to about 20 and 50\,per cent, 
respectively, are thus
expected. 

Note that the Chapman et al. (2003) low-redshift {\it LT} relation shown by the 
dotted lines in Fig.\,8 is consistent with estimates of 
the temperature of the bulk of the submm-selected galaxy 
population derived by comparing their multiwavelength properties 
(Blain et al.\, 1999; Trentham, Blain \& Goldader 1999). Using 
the $T$--$\alpha$--$\beta$ SED description (equation 1), these temperatures 
lie close to 40\,K at luminosities  
of several $10^{12}$\,L$_\odot$, assuming a high redshift ($z>1$).  

The SEDs discussed earlier, for NGC\,958, Arp\,220 and APM\,08279+5255,  
can be analyzed in the same way as the data for SMM\,J14011+0252 
shown in Fig.\,8.\footnote{ 
SMM\,J14011+0252 seems to set an unfortunate precedent for studies of 
submm galaxies: 
it is unusually bright in the optical range, and it shows no sign of 
the presence of an AGN, unlike most other 
identified galaxies (Smail et al.\ 2002). 
It lies squarely on 
the radio--far-IR correlation. Although it is relatively 
easy to study, SMM\,J14011+0252 is perhaps unrepresentative of 
submm galaxies as a whole.} The resulting 
tracks in the {\it LT} diagram are shown in Fig.\,9. 
The {\it LT} curves for 
NGC\,958 and Arp 220 intersect with the low-redshift Chapman et al. {\it LT} 
relation at temperatures of approximately 28 and 38\,K respectively. These 
results are very close to their fitted temperatures, taking into account 
their redshifts, which are 29 and 38\,K respectively in model\,1 (Table\,1). 
Hence, in the absence of redshift information, a photometric redshift 
derived for both of these galaxies would be reliable, although the 
uncertainty would be dominated by the {\it LT} relation.   

However, the curves intersect for APM\,08279+5255 at $T \simeq 34$\,K, 
nowhere near its true temperature 
$T \simeq 80$\,K (in model\,1), even  
after correcting for magnification by an assumed factor of 50.
This is at least a warning 
that some galaxies would have a very discrepant 
photometric redshifts derived from far-IR and submm data 
using this technique. This galaxy is significantly 
hotter than other high-redshift dusty galaxies, but it is certainly an 
interesting object, and owing to its great luminosity, one that could be 
found quite easily in future far-IR and submm-wave surveys. 
The observational 
errors for all three of these brighter galaxies 
are much smaller than for SMM\,J14011+0252, and so the  
discrepancies in their photometric redshifts definitely reflect a dispersion in 
the {\it LT} relation rather than errors in the data points.

In order to estimate reliable photometric redshifts  
from submm, far/mid-IR and radio observations it is necessary to 
be certain of the nature of and scatter in the {\it LT} relation. This can be 
investigated using a variety of samples of IR-luminous galaxies with 
known redshifts. It is also important to determine whether the relation 
evolves with redshift. If it does, then this could provide insight into 
the astrophysics of dusty galaxies, in addition to important information for 
finding photometric redshifts. 

\begin{figure}
\begin{center}
\epsfig{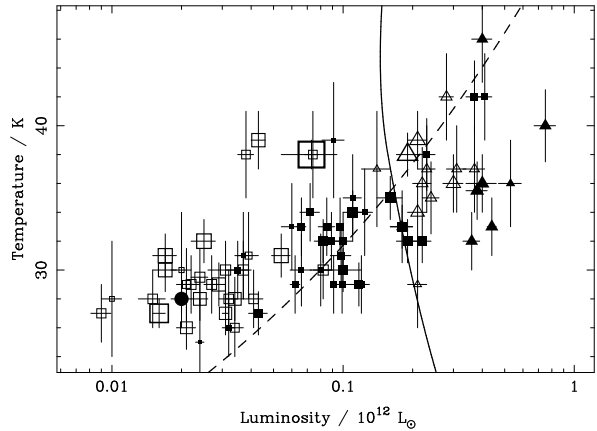}
\end{center}
\caption{The {\it LT} values derived for 83 {\it IRAS} BGS galaxies observed 
by SCUBA (Dunne et al.\ 
2000), 
fitted using the $T$--$\alpha$--$\beta$ SED description.
The different symbols represent different redshifts: $z<0.01$, open 
square; $0.01 \le z<0.02$, filled square; $0.02 \le z < 0.03$, empty 
triangle; $0.03 \le z < 0.04$, filled triangle; $0.04 \le z < 0.05$, 
empty circle; and $z \ge 0.05$, filled circle. Larger symbols 
represent more accurate results. 
The overplotted solid and dashed lines trace the
loci of a 0.5-Jy 60-$\mu$m source and a 60-mJy 850-$\mu$m source, respectively,
both at $z=0.02$. They 
show the direction in which 
observational selection effects could truncate the distribution of 
points.
The dashed 850-$\mu$m 
curve runs parallel to the distribution of the data points, and so there 
might be a selection effect against detecting hot sources 
in the sample. However, 
almost all the targeted {\it IRAS} sources were detected at 
850\,$\mu$m, and so the lack of sources at the top left of the field 
probably reflects a genuine absence of hot, low-luminosity galaxies in 
the {\it IRAS} sample.} 
\end{figure} 

\begin{figure}
\begin{center}
\epsfig{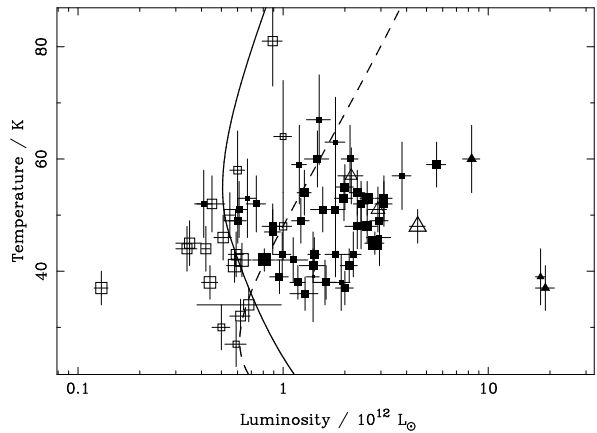}
\end{center}
\caption{The {\it LT} relation of 72 galaxies in both the {\it IRAS} Faint 
Source Catalog
and the VLA-FIRST radio survey catalogue 
(Stanford et al.
2000),  
fitted using the $T$--$\alpha$--$\beta$ SED description.
The different symbols represent different redshifts: $z<0.1$, open
square; $0.1 \le z<0.2$, filled square; $0.2 \le z < 0.3$, empty
triangle; $0.3 \le z < 0.4$, filled triangle; $0.4 \le z < 0.5$,
empty circle; and $z \ge 0.5$, filled circle. 
Better fitting data are plotted using
a larger symbol. The solid and dashed overplotted lines show the direction 
in which
observational selection effects would be important, and trace the
loci of a 100-mJy 60-$\mu$m source and a 1-mJy 1.4-GHz source, respectively, 
both at $z=0.25$. 
Both curves 
cut directly across the cloud of points, and so the lack of sources away 
from the cloud of points at luminosities $L > 10^{12}$\,L$_\odot$ 
is not likely to be due to selection effects. 
}
\end{figure} 

\begin{figure}
\begin{center}
\epsfig{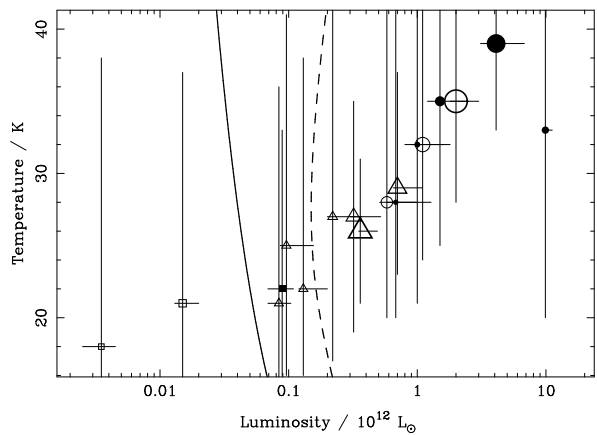}
\end{center}
\caption{The {\it LT} relation for 18 galaxies 
detected at radio wavelengths using MERLIN and in the mid-IR 
using {\it ISO} in 
the HDF (Garrett\ 2002), fitted using 
the $T$--$\alpha$--$\beta$ SED description (equation 1).
The different symbols represent different redshifts: $z<0.2$, open
square; $0.2 \le z<0.4$, filled square; $0.4 \le z < 0.6$, empty
triangle; $0.6 \le z < 0.8$, filled triangle; $0.8 \le z < 1$,
empty circle; and $z \ge 1$, filled circle. Larger symbols
represent more accurate results. 
Note that the temperatures are derived by 
assuming that the far-IR--radio correlation holds, and their errors 
are much larger than those for the other samples shown in Figs\,10 
and 11. 
The solid and dashed overplotted lines show the direction in which
observational selection effects would be important, and 
trace the
loci of a 0.5-mJy 15-$\mu$m source and a 50-$\mu$Jy 1.4-GHz source 
respectively, both 
at $z=0.5$. 
Both lines 
cut directly across the points, and so the lack of sources away 
from the trend is not likely to arise from selection effects. 
}
\end{figure}

\subsection{Observed {\it LT} relations} 

The accurate determination of the temperature and luminosity of dusty 
galaxies with known redshifts using a three-parameter $T$--$\alpha$--$\beta$ 
SED description was illustrated above.  
In order to determine a temperature reliably, 
the redshift must be known, and flux density data must be 
available at frequencies both above and below the peak of the SED. Radio 
data can be used as a proxy for mid-/far-IR data if the galaxies can be 
assumed to lie on the far-IR--radio correlation. The 
number of galaxies for which all the required far-IR/radio and 
submm information is available is relatively small. We now derive  
{\it LT} relations from three different samples of galaxies to 
investigate its properties from the limited existing data. 

\subsubsection{Low-redshift {\it IRAS} galaxies} 

We have already discussed the low-redshift {\it LT} relation derived from a 
large sample of {\it IRAS} galaxies by Chapman et al.\ (2003): see 
the dotted lines in Figs\,8 and 9. However, these results are 
based on radio and 
far-IR data alone. It would be very useful to include submm 
data at intermediate wavelengths in order to be sure of the form of 
the SED. 

A much smaller number of low-redshift galaxies in the {\it IRAS} 
catalogue were 
observed at 
submm wavelengths in the SLUGS survey 
by Dunne et al.\ (2000) and Dunne \& Eales (2001).
The results provide important information about both the 
local, low-luminosity {\it LT} relation and galaxy SEDs. 
Many of these 
galaxies also have radio data from the NED database, and we have 
exploited this information, assuming the far-IR--radio 
correlation to improve the accuracy of the derived values of bolometric 
luminosity and 
temperature for these galaxies as compared with the SLUGS 
values. The resulting SEDs are thus obtained by combining radio data with 
the flux densities at 850, 100 and 60\,$\mu$m considered by Dunne et al.\ 
(2000). 
The resulting quantities for 
83 galaxies in the SLUGS sample 
are shown in 
Fig.\,10, along with two lines that trace the luminosity and temperature of 
a galaxy that is required to generate the typical flux density of a 
galaxy in the sample at a 
typical redshift for the sample. These curves provide an indication of 
the possible role of  
selection effects in 
limiting the extent of the scatter in the derived 
{\it LT} relation. If the lines lie 
parallel to a correlation in the plotted points, then  
some of the correlation could be due to selection effects acting to reduce the 
intrinsic scatter, by removing galaxies from the sample on 
the low-luminosity side of the line. This is not the only way in which 
selection effects could modify the observed or inferred scatter in an 
{\it LT} relation; however, it provides a direct indication of whether or not 
selection effects are likely to be significant. 

For the SLUGS data, the role of selection effects 
is unlikely to be very significant, as almost all of the targeted {\it IRAS}
galaxies were detected at submm wavelengths (Dunne et al.\ 2000). 
Note that the luminosities 
and temperatures of the Milky 
Way (Reach et al. 1995; $L \sim 3 \times 10^{10}$\,L$_\odot$; $T \simeq 17$\,K) 
and NGC\,891 (Alton et al.\ 1998; $L \simeq 5.3 \times 10^9$\,L$_\odot$; 
$T \simeq 20 \pm 2$\,K) lie at luminosities slightly below the 
cloud of data points, whereas the low-luminosity, hot starburst galaxy M82 
($L \simeq 2.7 \times 10^{10}$\,L$_\odot$; $T = 42 \pm 2$\,K) is significantly 
offset above the cloud. The apparent dispersion in the 
{\it LT} relation shown in Fig.\,10 may thus be less than the true dispersion. The 
scatter in the temperatures derived for SLUGS galaxies  
is comparable to the scatter determined from the SEDs of low-redshift 
{\it IRAS} galaxies shown in Fig.\,8 (Chapman et al.\ 2003). 

\subsubsection{VLA--{\it IRAS} galaxies at moderate redshifts} 

At higher redshifts, {\it IRAS} flux 
densities are available only for the most luminous galaxies. The fluxes 
from the {\it IRAS} Faint Source Catalog (FSC) were correlated 
with data from the wide-field VLA-FIRST radio survey (Becker, 
White \& Helfand 1995) to provide information on the {\it LT} relation for 
a large sample of galaxies with bolometric luminosities 
greater than $\sim 10^{12}\,$L$_\odot$ by 
Stanford et al.\ (2000). The {\it LT} values for these galaxies were 
calculated assuming that the far-IR--radio correlation holds. 
The {\it LT} values for the 72 
galaxies from this sample with reliable redshifts, which are not fitted by 
extremely cold temperatures ($T<5$\,K) 
and thus radio-loud,  
are plotted in Fig.\,11. About 25 further galaxies from the sample of 
Stanford et al. 
fall into the 
very cold category, thus indicating a likely 
$\sim 25$\,per cent contamination fraction  
in the sample from radio-loud AGN that emit a radio flux density greater 
than that expected from the observed low-redshift far-IR--radio correlation. 
The tracks of the lines through the data 
confirm that neither radio nor IR selection effects should 
severely bias the {\it LT} relation from the sample of Stanford et al., which  
lies on a slightly hotter track, and is scattered by a greater amount to 
higher temperature as compared with the SLUGS sample. 
This suggests that hotter, perhaps more AGN-rich galaxies,
are represented in this more luminous, partially radio-selected 
sample. 
A similar sample of about 40 luminous southern galaxies, 
with radio images from the Molonglo telescope and redshifts from 
the 2dF multi-object spectrograph has recently been compiled by Sadler 
et al.\ (2002), and could be analyzed in a similar way. 
Our understanding of the {\it LT} relation would be much 
improved if submm fluxes could be determined for these galaxies 
with known moderate redshifts and radio and far-IR flux densities. 
These observations would both provide tighter 
constraints on their positions in the {\it LT} plane, and 
offer the possibility to search for evolution in the {\it LT} relation 
(Chapman et al. 2003). 

\subsubsection{Faint radio and mid-IR selected galaxies in the 
Hubble Deep Field} 

Another sample useful for investigating the {\it LT} relation is 
the 18 $z \sim 1$ faint radio galaxies detected at 15\,$\mu$m using 
{\it ISO} in the Hubble Deep Field 
(HDF; Garrett 2002). These galaxies are sampled at mid-IR 
wavelengths much shorter than the 
peak of their SEDs, but the radio data provides a proxy for the  
60--100-$\mu$m emission close to the peak, assuming that the 
far-IR--radio correlation  
holds. A coarse upper limit to the 60-$\mu$m flux densities of the 
galaxies is 
provided by an XSCANPI 
analysis\footnote{http://www.ipac.caltech.edu/ipac/services/xscanpi.html} 
of the relevant 
{\it IRAS} scans. A constraint on the 60-$\mu$m flux densities of 
these galaxies is essential for fitting  
temperatures to the radio--mid-IR data. 
The inferred {\it LT} relation 
is shown in Fig.\,12:
it exhibits a remarkably narrow 
dispersion, but the errors on the data points are large. 
Galaxies at the mid-IR wavelengths probed have spectral features 
associated 
with emission from PAH molecules, and so 
this tight correlation is all the more surprising. A link between the 
intensity of PAH emission and the bolometric luminosity of a galaxy has 
been commented on for a sample of only five galaxies by Haas, Klass \& Bianchi 
(2002). An apparent link between 15\,$\mu$m emission and bolometric luminosity 
for a larger sample of low-redshift galaxies is discussed in Sections 5.3 and 
5.5 of Dale \& Helou (2002). The tracks of 
typical galaxies in the sample overplotted 
on Fig.\,11 show that selection effects are unlikely to be responsible for 
the tight correlation. 

It seems unlikely that the mid-IR flux density of a galaxy could 
be better correlated with its temperature and luminosity than data obtained 
closer to the peak of the SED. This would require some underlying correlation 
between the slope of the mid-IR SED and the total luminosity that is 
not apparent in existing datasets. Much larger samples from 
{\it SIRTF} from 2003 will hopefully resolve this question. 

\subsubsection{Submm-selected high-redshift galaxies} 

A few distant submm- and 
far-IR-selected
galaxies (Ivison et al.\ 1998, 2000, 2001; Frayer et al.\ 1998, 1999; 
Chapman et al.\ 2002a) also have
redshifts, and many more have recently been found using deep radio-selected 
galaxies (Chapman et al.\ 2002b). 
Initial results indicate that it is likely that the values of 
temperature and luminosity derived for  
high-redshift submm-selected dusty galaxies
display a similar scatter to that shown in Fig.\,11. These samples are 
destined to grow in size, and will provide the ultimate test of the 
high-redshift {\it LT} relation and the 
reliability of far-IR photometric redshifts for high-redshift galaxies.
This sample is extremely useful, as it consists of 
the target population for which photometric redshifts are
sought at far-IR and submm wavelengths. Inferred values of luminosity and 
temperature for high-redshift dusty galaxies with known 
redshifts can be found in 
Chapman et al.\ (2002b). 

\subsection{{\it LT} relations from a combination of samples} 

The combined {\it LT} relation for all three samples is compared in Fig.\,13, 
segregated between samples by the plotting symbol. The 
{\it LT} relations estimated for low-redshift {\it IRAS} galaxies by Chapman 
et al.\ (2003), based on 60--100\,$\mu$m colours and the SED templates 
of Dale et al.\ (2001), and for both merging and quiescent galaxies at 
low redshifts by 
Barnard \& Blain (2003) are represented by the lines. 
Note that the temperature inferred 
by Chapman et al.\ (2003) is $T_{\rm min}$ as 
defined in equation (4), and 
so is lower by 10--20\,per cent as compared with the definitions used here, 
reducing the difference between the solid line that represents the 
Chapman et al. {\it LT} relation and the dashed line representing 
the {\it LT} relation for luminous merging galaxies. 

The dispersion in the {\it LT} relation exceeds 25\,per cent (0.1\,dek) at all 
luminosities. 
These results demonstrate that the dispersion in the 
{\it LT} relation within individual 
samples is less than the dispersion between samples. 
The Stanford et al. sample 
is more likely to be significantly affected by 
radio emission from AGN, which could introduce additional scatter,  
causing an overestimate of luminosity and an underestimate of temperature. 
High-luminosity sources that would be selected in submm-wave 
surveys are thus observed to 
have dust temperatures that range between 25 and 60\,K. The median 
temperature 
and its RMS scatter in temperature are  
about $45 \pm 10$\,K, a dispersion of approximately 0.1\,dek.  

Galaxies with low dust temperatures are difficult to select 
in the far-IR surveys from which these {\it LT} results are drawn, and so there 
may not necessarily be a lack of very 
luminous cool galaxies, despite the avoidance of points at the lower 
right corner of Fig.\,13. By contrast, there appears to be a real lack 
of hot, low-luminosity galaxies from the SLUGS and Stanford et al. 
{\it IRAS}-selected galaxies. An exception is one of the closest, 
brightest {\it IRAS} galaxies, M82. Despite its low luminosity of 
only $2.7 \times 10^{10}$\,L$_\odot$, M82 has a temperature of $42 \pm 2$\,K, 
in the empty upper left region of the figure. Such hot, dwarf 
galaxies are difficult to find in the {\it IRAS} survey, on the grounds 
of the limited survey volume for low-luminosity galaxies; however, much deeper 
surveys using {\it SIRTF} may find that this region of the {\it LT} plane is 
more thoroughly populated with moderate-redshift galaxies.

The same points are replotted in Fig.\,14, segregated by the plotted symbol  
in redshift rather than from sample to sample. 
There is a natural tendency for more 
luminous galaxies to be selected at greater redshifts. As the population 
of dusty galaxies is known to evolve strongly in luminosity with 
redshift, this trend should be all the more apparent.  
However, at 
all redshifts the scatter in the {\it LT} relation from sample to sample appears to 
span the full range of the scatter seen in the combined 
population. The dispersion 
thus appears to reflect 
a real spread in the physical properties of galaxies present 
at all redshifts, and not to be due solely to 
a systematic evolution in a tightly dispersed {\it LT} relation with redshift.
There is no strong evidence in samples of {\it IRAS}-selected galaxies 
for the {\it LT} relation to evolve with redshift over the approximate range 
$0<z<0.3$ (Chapman et al.\ 2003). 

The significant dispersion in the {\it LT} relation indicates 
that there are limited opportunities for discriminating between hot--distant  
and cool--closer 
galaxies on the grounds of their far-IR/submm colours even with an assumed {\it LT}
relation. With an {\it LT} dispersion of at least 25\,per cent (Fig.\,13), the 
fractional accuracy of a photometric redshift determination is never likely 
to be better than 30\,per cent. Current evidence thus suggests that 
spectroscopic redshifts will remain essential to 
interpret the nature of submm-selected galaxies, unless increased 
sample sizes reveal that 
the existing samples of high-redshift 
dusty galaxies are scattered 
by a larger amount in the {\it LT} plane 
than the true distribution, which is currently 
well determined only at low redshifts (Chapman et al.\ 2003). We think that 
this is unlikely, and rather that the observed 
scatter in the distribution will remain 
the same or grow larger as more information becomes available. 
Larger samples will provide a better description of the distribution of 
galaxies in the wings of the {\it LT}
distribution, and unearth further examples of rare galaxies like
M82 and APM\,08279+5255, that are not represented 
in the {\it LT} plane at the current sampling rate. 

\begin{figure}
\begin{center}
\epsfig{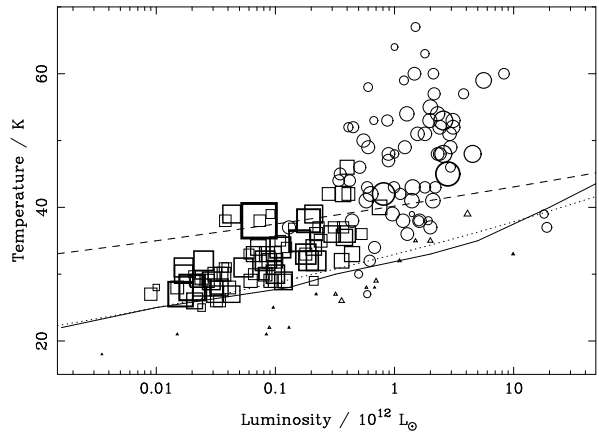}
\end{center}
\caption{The combined {\it LT} scatter for all galaxies shown in Figs\,9--11:  
Dunne et al.\ (2000), squares; 
Garrett (2002), lozenges; and Stanford et al.\ (2000),
circles. The better-fitting  
galaxies are represented by larger symbols. 
The overplotted lines show the results from other low-redshift 
{\it LT} investigations: the solid 
line shows the result of Chapman et al.\ (2003), and the 
dashed and dotted lines 
represent the results for merging and quiescent galaxies by 
Barnard \& Blain (2003) respectively. Galaxies appear to avoid 
the hot, low-luminosity region to the upper left of the 
figure, which is not likely to be empty due to systematic 
selection effects. 
}
\end{figure}

\begin{figure}
\begin{center}
\epsfig{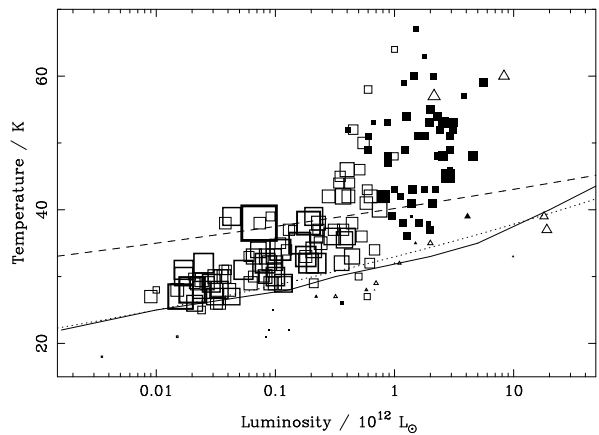}
\end{center}
\caption{The combined {\it LT} relation illustrated in Fig.\,12, this time 
as a function of redshift. 
Galaxies at $z \le 0.2$ are represented by empty squares, at 
$0.2<z \le 0.5$ by filled squares, at $0.5 < z \le 1$ by empty triangles, 
at $1 < z \le 2$ by filled triangles, at $2 < z \le 3$ by empty circles 
and at $z>3$ by filled circles. There is a natural trend to 
greater luminosities/temperatures at higher redshifts. In each redshift 
interval temperatures range over the full 
extent of the scatter in the diagram, and there is no evidence for a 
significant change in the 
{\it LT} relation with redshift (Chapman et al.\ 2003).
}
\end{figure}  

\section{Future SED measurements} 

The greatest change in our understanding of the properties of dusty 
galaxies will be brought with the launch of {\it SIRTF} in 2003 January.
With a 0.85-m aperture, and capable of diffraction-limited 
imaging in far-IR bands at effective wavelengths of 24, 70 and 160\,$\mu$m, 
{\it SIRTF} will provide key information on the SEDs of at least 
several million galaxies. Unlike galaxies detected by {\it IRAS}, 
these will extend well beyond a redshift of 1, and will be located 
to an accuracy of order 5\,arcsec. By combining the positions of 
galaxies detected by {\it SIRTF} with spectroscopic redshifts obtained 
as part of the 
Sloan Digital Sky Survey (SDSS), which is reasonably complete to 
$z \simeq 0.3$, it should be possible to obtain SED data for 
of order 10$^4$ galaxies with exact redshifts, and thus to determine 
the form of the {\it LT} relation discussed in Fig.\,14 in much more detail. 
At that point the accuracy of photometric redshifts based on far-IR 
colours combined with an {\it LT} relation can be assessed realistically, 
at least at low to moderate redshifts. 

The information can be extended out to higher redshifts by considering
optical photometric redshifts, derived reliably from spectral breaks in the 
SDSS data for galaxies too faint/distant to be targeted 
for SDSS spectroscopy. At the same time, many of the galaxies 
with known high redshifts shown in Fig.\,14, which already have some 
spectral information at far-IR/submm wavelengths, will be 
targets for {\it SIRTF}. More information will thus soon be available 
about their SEDs. 
By combining radio observations with several far-IR data points 
it should be possible to generate a more accurate {\it LT} 
relation, and to measure any changes in the far-IR--radio 
correlation with increasing redshift by 
the end of {\it SIRTF}'s mission in approximately 2008. 
The true potential for far-IR and submm-wave 
photometric redshifts can then be assessed securely.    
The reliability of the 
far-IR--radio correlation can in the meantime be investigated using 
data for any galaxy with a known redshift that has three accurate 
flux density 
measurements, one each at radio, submm and far-IR wavelengths. 

\section{Conclusions} 

We have discussed the description of 
the SEDs of dusty galaxies using four different models
that are appropriate to describe  
data that is available at present and is likely to be generated by 
forthcoming space missions. 
One of the parameters in each model always describes the peak frequency 
of the thermal dust SED (`temperature'), while two spectral 
indices describe the fraction of hot and cold dust: an `$\alpha$'
and `$\beta$' parameter respectively.
Observational data 
constrains these SED descriptions to within 10\,per 
cent accuracy across the full range of interesting wavelengths 
longer from about 20\,$\mu$m to deep in the radio waveband. It is 
important to be careful in interpreting the values of dust 
temperatures, emissivities and masses that are inferred with 
the values of real physical 
parameters. 

There is a huge degeneracy between temperature and redshift 
when fitting the SED of a distant galaxy. 
Assuming some link between the luminosity and SED  
allows this degeneracy to be broken, but a range of available 
information indicates that this relationship has a very 
considerable scatter, by up to a factor of 2. Without the  
knowledge that this {\it LT} relation 
has a scatter as narrow as the required 
accuracy of the photometric redshift, 
continuum far-IR/submm/radio 
photometric redshifts are almost useless for providing constraints from 
the data  
for an individual galaxy. 

In order to finally assess their usefulness  
it is essential to quantify the {\it LT}
relationship accurately, based on a large number of galaxies with 
known redshifts and well-sampled SEDs. This information is not 
available at present, but will be generated by 
{\it SIRTF}. Only if the true scatter in 
the {\it LT} relation turns out to be less than about 20\,per cent will the  
photometric redshift technique be
useful. 
Spectroscopic observations to fix the redshifts and 
SEDs of dusty galaxies remain essential to understand the population, and 
it is important to develop new types of spectroscopic instruments 
that can address these questions, for example wide-band detectors for 
multiple CO emission lines (Bradford et al. in preparation).

\section*{Acknowledgments}

This research has made use of the NASA/IPAC Extragalactic Database (NED)   
which is operated by the Jet Propulsion Laboratory, California Institute   
of Technology, under contract with the National Aeronautics and Space      
Administration. We thank the referee Danny Dale for a very rapid and 
helpful report.

\end{document}